\documentclass[journal]{IEEEtran}
\ifCLASSINFOpdf
\else
\fi
\usepackage[cmex10]{amsmath}
\usepackage{array}
\usepackage{url}
\usepackage{amsfonts}
\usepackage{amssymb}
\usepackage{xcolor}

\begin{document}

\newtheorem{theorem}{Theorem}
\newtheorem{cor}[theorem]{Corollary}
\newtheorem{definition}{Definition}
\newtheorem{example}{Example}
\newtheorem{lemma}[theorem]{Lemma}
\newtheorem{proposition}{Proposition}
\newtheorem{fact}{Fact}
\newtheorem{property}{Property}
\newcommand{\bra}[1]{\langle #1|}
\newcommand{\ket}[1]{|#1\rangle}
\newcommand{\braket}[3]{\langle #1|#2|#3\rangle}
%%inner product
\newcommand{\ip}[2]{\langle #1|#2\rangle}
%%outer product
\newcommand{\op}[2]{|#1\rangle \langle #2|}
\newcommand{\lin}[1]{\setft{L}\left(#1\right)}

\newcommand{\tr}{{\rm tr}}
\newcommand{\supp}{{\it supp}}
\newcommand{\sch}{{\it Sch}}

\def\complex{\mathbb{C}}
\newcommand {\E } {{\mathcal{E}}}
\newcommand {\F } {{\mathcal{F}}}
\def\X{\mathcal{X}}
\def\Y{\mathcal{Y}}
\newcommand {\diag } {{\rm diag}}

\title{Distinguishability of Quantum States by Positive Operator-Valued Measures with Positive Partial Transpose}

\author{Nengkun Yu,\ \ Runyao Duan and Mingsheng Ying\thanks{This work was partly supported by the National Natural Science Foundation
of China (Grant Nos. 61179030 and 60621062) and the Australian Research Council (Grant Nos. DP110103473, DP120103776, and FT120100449).

N. Yu is currently with the Institute for Quantum Computing and the Department of Combinatorics \& Optimization, University of Waterloo, Waterloo, and the Department of Mathematics and Statistics, University of Guelph, Guelph, ON, Canada. He was affiliated with the State Key Laboratory of Intelligent Technology and Systems, Tsinghua National Laboratory for Information Science and Technology, Department of Computer Science and Technology, Tsinghua University, Beijing 100084, China, and the Centre for Quantum Computation and Intelligent Systems, Faculty of Engineering and Information Technology, University of Technology, Sydney, NSW 2007, Australia while this work was completing.

R. Duan and M. Ying  are with the Centre for Quantum Computation and Intelligent Systems, Faculty of Engineering and Information Technology, University of Technology, Sydney, NSW 2007, Australia. They are also affiliated with the UTS-Tsinghua Joint Research Centre for Quantum Computation and Artificial Intelligence, the State Key Laboratory of Intelligent Technology and Systems, Tsinghua National Laboratory for Information Science and Technology, Department of Computer Science and Technology, Tsinghua University, Beijing 100084, China, and the UTS-AMSS Joint Research Laboratory for Quantum Computation and Quantum Information Processing, Academy of Mathematics and Systems Science, Chinese Academy of Sciences, Beijing 100190, China.

The material in this paper was presented in part as a long talk at the Asian Quantum Information Science Conference (AQIS12), Suzhou, China, August 2012.}}

\maketitle

\begin{abstract}
We study the distinguishability of bipartite quantum
states by Positive Operator-Valued Measures with positive partial
transpose (PPT POVMs). The contributions of this paper include: (1). We give a negative answer to an open problem of [M. Horodecki $et. al$, Phys. Rev.
Lett. 90, 047902(2003)] showing a limitation of their method for detecting nondistinguishability. (2). We show that a maximally entangled state
and its orthogonal complement, no matter how many copies
are supplied, can not be distinguished by PPT POVMs, even
unambiguously. This result is much stronger than the previous
known ones \cite{DUAN06,BAN11}. (3). We study the entanglement cost of distinguishing quantum states. It is proved that $\sqrt{2/3}\ket{00}+\sqrt{1/3}\ket{11}$ is sufficient and necessary for distinguishing three Bell states by PPT POVMs. An upper bound of entanglement cost of distinguishing a $d\otimes d$ pure state and its orthogonal complement is obtained for separable operations. Based on this bound, we are able to construct two orthogonal quantum
states which cannot be distinguished unambiguously by separable
POVMs, but finite copies would make them perfectly
distinguishable by LOCC. We further observe that a two-qubit maximally
entangled state is always enough for distinguishing a $d\otimes d$ pure state and its orthogonal complement by PPT POVMs, no matter the
value of $d$. In sharp contrast, an entangled state with Schmidt
number at least $d$ is always needed for distinguishing such two
states by separable POVMs. As an application, we show that
the entanglement cost of distinguishing a $d\otimes d$ maximally entangled
state and its orthogonal complement must be a maximally entangled state for $d=2$,
which implies that teleportation is optimal; and in general, it could be chosen as $\mathcal{O}(\frac{\log d}{d})$.
\end{abstract}


% \fi
\IEEEpeerreviewmaketitle

\section{Introduction}
One of the main goals of quantum information theory is to
understand the power and limitation of quantum operations
which can be implemented by local operations and classical
communication (LOCC). These are operations wherein two or
more physical distant parties retaining the ability of performing
arbitrary operations on the quantum system one part holds,
and the result of local operations can be \textquotedblleft communicated"
classically to another part. The class of LOCC operations
provides a natural setting to address intrinsic problems about quantum
nonlocality and entanglement.

Quantum information is nonlocal in the sense that local measurements on a multipartite quantum system,
prepared in one of many mutually orthogonal states, may not reveal in which state the system was
prepared. In the widely studied bipartite case, the scenario is that one of known
orthogonal quantum states is shared by two parties, says Alice
and Bob, and their goal is to identify which of the state it is; see Ref.
\cite{YDY11b,WSHV00,GKRS+01,GKRS04,BDFM+99,BDMS+99,BGK11,WAT05,YDY11a,HSSH03,DFJY07,DFXY09,NAT05,ALE12,HMM+06} as a very incomplete list. In some situations Alice and Bob are able to accomplish
this task without error, but in others they are not. For example,
Walgate \textit{et. al} \cite{WSHV00} proved that any two orthogonal pure states,
no matter entangled or not, are locally distinguishable with
no error. On the other hand, examples of orthogonal product states that
can not be distinguished by LOCC protocols are presented, for instance, a two-qutrit
orthonormal pure product basis \cite{BDFM+99} and any set
of states forming an unextendible product bases \cite{BDMS+99}.
Horodecki \textit{et.al} \cite{HSSH03} discovered a phenomenon of \textquotedblleft more nonlocality with less entanglement''. These examples demonstrate
that entanglement is not always decisive feature of locally
distinguishability. It is thus necessary to further clarify
the role of entanglement in the local distinguishability
in differenct circumstances. Considerable efforts have been
devoted to the local discrimination of maximally entangled
states. Large set of maximally entangled states cannot be
distinguished locally: if Alice and Bob's system
are $d$-dimensional spaces, then it is impossible for them to
distinguish $d+1$ or more maximally entangled states perfectly
\cite{GKRS+01,GKRS04,YDY11b,NAT05,ALE12,HMM+06,OH03}. It is proved that three orthogonal
two-qutrit maximally entangled states are always locally distinguishable
\cite{NAT05}. We showed that $d+1$ is not a tight lower
bound for the number of locally indistinguishable maximally
entangled states by presenting four locally indistinguishable orthogonal ququad-ququad
maximally entangled states  \cite{YDY11b}. To circumvent the difficulty of
proving local indistinguishability, our approach is to show indistinguishability by PPT POVMs, and local indistinguishability
automatically follows since LOCC POVMs is a proper subset
of PPT POVMs. The advantage of this approach is that
the set of PPT POVMs enjoys a mathematical structure much simpler
than that of LOCC POVMs due to the complete characterization
of PPT condition by semi-definite programming. After our
work, several examples of $d$ PPT indistinguishable $d\otimes d$
maximally entangled states are found by using semi-definite
programming \cite{ALE12}.

The notion of PPT plays a significant role in quantum
information theory. First, it has been used to provide some
convenient criterion for the separability of quantum states,
which is one of the central topics in quantum information
theory and has been extensively studied in the last two
decades. Peres \cite{PER96} proved that any separable state should obey the
PPT criterion. Horodecki \textit{et.al} \cite{HHH96} established a connection
between separability and positive maps acting on operators
and used it to prove that PPT criterion is also sufficient
for the separability of $2\otimes 2$ or $2\otimes 3$ states. They also
showed that if a mixed state can be distilled to the singlet
form, it must violate the PPT criterion \cite{HHH98}. It has been
conjectured that NPT bound entangled state does exist, and
this remains one of the most important open problems in
quantum information theory. Also PPT operations have been
used to study the problem of entanglement distillation and
pure state transformation \cite{ISH04,MW08,RAI01}. Ishizaka \cite{ISH04} showed that
bipartite pure entangled states can be transformed to arbitrary
pure states by stochastic PPT operations.

The first purpose of this paper is to further study the strength and
limitation of PPT POVMs by considering the distinguishability
of quantum states under PPT POVMs. In other words, given a known set of mutually orthogonal states, we
may wish to know whether it is possible for the parties to perfectly
distinguish the state; that is, given any one of the states in the
set and by using PPT POVMs, can they with certainty determine
which state they were given? We study this problem by starting with an observation that some results of state discrimination by separable POVMs in \cite{DFXY09} can be directly generalized to the case by PPT POVMs. More precisely, we give a necessary and sufficient condition for the distinguishability
of a set of quantum states by
PPT POVMs. Leveraging this condition,
the problem of distinguishing $(D-1)$ bipartite pure states by
PPT POVMs is showed to be equivalent to that of distinguishing them by
separable POVMs, where $D$ is the total dimension of the state space
under consideration. We show that the orthogonal
complement of a bipartite pure state has a
PPT distinguishable basis if and only if the Schmidt number of this state
is less than 3.

In recent years, entanglement has already been shown to be a valuable resource,
allowing remote parties to communicate in ways that
were previously not thought possible. For instance, any set of orthogonal states that cannot be distinguished by LOCC
alone can nonetheless always be distinguished by LOCC if
the parties share enough entanglement. The reason is that any global operation can be implemented by LOCC with help of entanglement by using teleportation.
It is of fundamental interest to understand the role of entanglement resource plays in certain tasks. That is, how much entanglement is needed to reach the goal which is impossible to be accomplished without entanglement.

Our second purpose of this work is to study the problem of entanglement cost of distinguishing those PPT indistinguishable quantum states by PPT POVMs. The motivation of this part is from two side: The first is to approximate the entanglement cost of state discrimination by LOCC. Though the structure of LOCC POVMs are mathematically complicated, separable POVMs is believed as a good approximation of LOCC protocols for many cases, and the entanglement cost by PPT POVMs is a lower bound of that by separable POVMs. On the other hand, one can learn the difference between separable POVMs and PPT POVMs by comparing the different costs. Specifically, we study the entanglement cost of distinguishing two well-known examples of PPT indistinguishable states. The first example is to distinguish three Bell states. This example is interesting because Bell states play very important role in quantum information theory. The second example is to distinguish a pure state and its orthogonal complement, there are the reasons of studying this example: One is it reveals some major differences between the distinguishability of pure states and of mixed states. The other reason is that this simple example shows the sharp difference of distinguishability power between PPT POVMs and separable POVMs.

The major contributions of this paper include:
\begin{enumerate}
\item We solve the open problem proposed in \cite{HSSH03} based on our previous results of \cite{YDY11b}. In particular, we show that the HSSH method presented in \cite{HSSH03} is not a \textquotedblleft if and only if'' criterion for checking local distinguishability; more precisely, the indistinguishability of the ququad-ququad maximally entangled states considered in \cite{YDY11b} cannot be detected by the HSSH method.

\item By employing the technique introduced in \cite{YDY11b}, we show that a
   maximally entangled state and its orthogonal complement, no
   matter how many copies are supplied, can not be distinguished
   by PPT POVMs, even unambiguously. This is much stronger
   than the previous known results of \cite{DUAN06,BAN11}.

\item We study the entanglement cost of distinguishing quantum states. This
   problem is completely solved for the case of three Bell states: it is proved that
   $\sqrt{2/3}\ket{00}+\sqrt{1/3}\ket{11}$ is sufficient and necessary for distinguishing three
   Bell states by PPT POVMs. Then we consider how much entanglement is needed for distinguishing a $d\otimes d$ pure state and its orthogonal complement, and an upper bound of entanglement cost is obtained for separable operations. Based on this bound, we are able to construct two orthogonal quantum
   states which cannot be distinguished unambiguously by separable
   POVMs, but finite copies would make them perfectly
   distinguishable by LOCC. Furthermore, the entanglement cost for distinguishing a $d\otimes d$ pure state and its orthogonal complement by PPT POVMs is studied, and we show that a two-qubit
   maximally entangled state is always enough, no
   matter the value of $d$. In sharp contrast, an entangled state with
   Schmidt rank $d$ is always required for distinguishing such two
   states by separable POVMs. An interesting case is to distinguish a $d\otimes d$ maximally entangled state and its
   orthogonal complement. We show that for $d = 2$,
   the entanglement cost must be maximally entangled state, which can be
   interpreted as the optimality of teleportation. However, for sufficiently large $d$, the entanglement cost could be chosen
   arbitrarily close to 0.\end{enumerate}

%-----------------------------------------------------------------------------%
\section{Notations and Preliminaries} \label{sec:background}
%-----------------------------------------------------------------------------%

We first recall some notations of entanglement and preliminaries about state discrimination by LOCC POVMs, separable POVMs and PPT POVMs. Then we give a necessary and sufficient
condition for the distinguishability of a set of quantum states
by PPT POVMs. It should be pointed out that this condition is simply derived from a similar condition for  the distinguishability by separable POVMs provided in \cite{DFXY09}. Some applications of this condition can also be obtained by directly employing the condition of \cite{DFXY09}. For the reader's convenience, a detailed proof of this condition is included.

%-----------------------------------------------------------------------------%
\subsection{Basic linear algebra}
%-----------------------------------------------------------------------------%
In this paper the term {\it complex Euclidean space} refers to any
finite dimensional inner product space over the complex numbers.
Let $\X$ and $\Y$ be arbitrary
complex Euclidean spaces, and $\dim \X$ and $\dim\Y$ denote the dimensions of $\X$ and $\Y$, respectively. A pure quantum state of $\X$ is just a normalized vector $\ket{\Psi}\in \X$.

The space of (linear) operators mapping $\X$ to $\Y$ is denoted by
$\mathcal{L}(\X,\Y)$, while $\mathcal{L}(\X)$ is the shorthand for $\mathcal{L}(\X,\X)$. $I_{\X}$ is used to denote the identity operator on
$\X$.
The adjoint (or Hermitian transpose) of $A\in\mathcal{L}(\X,\X)$ is denoted by
$A^{\dag}$.
The notation $A\geq 0$ means that $A$ is positive semidefinite,
and more generally $A\geq B$ means that $A - B$ is positive semidefinite. $|A|=\sqrt{A^\dag A}$ is used to denote the positive square root of $A^{\dag}A$, i.e., $|A|=\sqrt{A^{\dag}A}$.

A general quantum state is
characterized by its density operator $\rho\in \mathcal{L}(\X)$, which is a positive
semi-definite operator with trace one on $\X$. The density
operator of a pure state $\ket{\psi}$ is simply the projector
$\psi:=\op{\psi}{\psi}$. The support of $\rho$, denoted by $\supp(\rho)$,
is the vector space spanned by the eigenvectors of $\rho$ with
positive eigenvalues. Alternatively, suppose $\rho$ can be
decomposed into a convex combination of pure states, say,
\begin{equation}\label{essemble}
\rho=\sum_{k=1}^n p_k\op{\psi_k}{\psi_k},
\end{equation}
where $0<p_k\leq 1$ and $\sum_{k=1}^n p_k=1$. Then
$\supp(\rho)=span\{\ket{\psi_k}:1\leq k\leq n\}$.

The Schmidt number of a bipartite state $\ket{\psi}\in \X\otimes\Y$ is defined as the minimum $k$ such that $\ket{\psi}=\sum_{i=0}^{k-1} \ket{\alpha_i}\ket{\beta_i}$ with unnormalized $\ket{\alpha_i}\in\X$ and $\ket{\beta_i}\in\Y$.
A pure state $\ket{\psi}\in \X\otimes\Y$ is called maximally entangled if $\ket{\psi}=\frac{1}{\sqrt{d}}\sum_{j=0}^{d-1}\ket{j}_X\ket{j}_Y$, where $\ket{j}_X$ and $\ket{j}_Y$ are orthonormal basis of $\X$ and $\Y$, respectively.
A bipartite mixed state $\rho\in\mathcal{L}(\X\otimes \Y)$ is said to be separable if in its decomposition of form (\ref{essemble}) all $\ket{\psi_k}$ can be chosen as
product states.

\begin{lemma}\label{invariant}\upshape
Let $\rho_1=\Phi\in\mathcal{L}(\X\otimes\Y)$ and $\rho_2=(I_{\X\otimes\Y}-\Phi)/(d^2-1)$, where $\ket{\Phi}=\frac{1}{\sqrt{d}}\sum_{j=0}^{d-1}\ket{j}_X\ket{j}_Y$ is maximally entangled state on $\X\otimes\Y$ with $d=\dim\X=\dim\Y$ and $\ket{j}_X$ and $\ket{j}_Y$ are computational basis of $\X$ and $\Y$, respectively. Then for any unitary $V\in\mathcal{L}(\X)$, we have
\[
(V\otimes V^*)\rho_k(V\otimes V^*)^{\dag}=\rho_k, k=1,2.
\] Moreover, for any $N\in \mathcal{L}(\X\otimes\Y)$, we have
\begin{eqnarray*}
\int\limits_{V} (V\otimes V^*)N(V\otimes V^*)^{\dag}dV=a\rho_1+b\rho_2,
\end{eqnarray*}
for some $a,b\in \complex$, where $V$ ranges over all unitaries in $\mathcal{L}(\X)$.
\end{lemma}
The validity of the above lemma can be verified by direct calculation.

The following lemma from \cite{DFXY09} is useful in the rest of this section.
\begin{lemma}\label{tech}\upshape
For $E\in\mathcal{L}(\X)$ such that $0\leq E\leq
I_{\X}$, and a density matrix $\rho$ on
$\X$, $\tr(E\rho)=1$ if and only if $E-P\geq 0$, where
$P$ is the projector on the support of $\rho$.
\end{lemma}
%-----------------------------------------------------------------------------%
\subsection{PPT distinguishability}
\label{sec:ppt}
%-----------------------------------------------------------------------------%
A nonzero positive semi-definite operator $E\in \mathcal{L}(\X\otimes \Y)$ is
said to be a PPT operator (or simply PPT) if $E^{\Gamma_{\X}}\geq 0$, where ${\Gamma_{\X}}$ means the partial transpose with respect to the party
$\X$, i.e.,
\begin{equation}
(\op{ij}{kl})^{\Gamma_{\X}}=\op{kj}{il}.
 \end{equation}
For simplicity, in what follows the subscript $\X$ of $\Gamma_{\X}$ will be omitted and $\Gamma$ is used instead of $\Gamma_{\X}$.

A Positive Operator-Valued Measure (POVM) on  $\X$ with $n$ outcomes is  an $n-$tuple of matrices, $(\Pi_k)_{k=1}^{n}$, where $\Pi_k\in\mathcal{L}(\X)$ with $\Pi_k\geq 0$ and $\sum_k \Pi_k=I_{\X}$.

Let $(\Pi_k)_{k=1}^{n}$ be a POVM acting on a bipartite system $\X\otimes\Y$. It is said to be a separable (SEP) POVM if $\Pi_k/(\tr(\Pi_k))$ is a
separable quantum mixed state for all $k$. It is said to be a PPT POVM if each $\Pi_k$ is PPT. It is known that any POVM that can be realized by means of an LOCC protocol is a PPT POVM. Moreover, we have
\begin{equation*}
LOCC~POVMs\subset SEP~POVMs \subset PPT~POVMs.
\end{equation*}

Let $\mathcal{S}=\{\rho_1,\cdots, \rho_{n}\}$ be a collection of $n$
quantum states. We say that $\mathcal{S}$ is perfectly
distinguishable by PPT (resp. SEP/LOCC) measurements if there is a PPT (resp. SEP/LOCC)
POVM $(\Pi_k)_{k=1}^{n}$ such that
\begin{equation}
\tr(\Pi_k\rho_j)=\delta_{k,j}
\end{equation}
for any $1\leq k,j\leq n$.

We say that $\mathcal{S}$ is unambiguously distinguishable by PPT (resp. SEP/LOCC) measurements if there is a PPT (resp. SEP/LOCC)
POVM $(\Pi_k)_{k=0}^{n}$ such that
\begin{equation}
\tr(\Pi_k\rho_j)=p_k\delta_{k,j}
\end{equation}
with some positive $p_k$ for any $1\leq k,j\leq n$.

It is obvious that unambiguous distinguishability is less constrained than ``normal" distinguishability.
\subsection{Distinguishability of quantum states by PPT POVMs}

It would be desirable to know when a collection of quantum
states is perfectly distinguishable by PPT POVMs. Generally,
orthogonality is not sufficient for the existence of a PPT
POVM discrimination. Noting the connection between separable and PPT, a rather simple necessary
and sufficient condition can be obtained by directly rewriting the proof of Theorem 1 in \cite{DFXY09}.
\begin{theorem}\label{exactdis}\upshape
Let $\mathcal{S}=\{\rho_1,\cdots,\rho_n\}$ be a collection of
orthogonal quantum states of $\X\otimes\Y$. Then $\mathcal{S}$ is
perfectly distinguishable by PPT POVMs if and only if
there exist $n$ positive semi-definite operators $\{E_1,\cdots,
E_n\}$ such that $P_k+E_k$ is PPT for each $1\leq k\leq n$,
and $\sum_{k=1}^n E_k=P_0$, where $P_k$ is the projector on
$\supp(\rho_k)$, and $P_0=I_{\mathcal{H}}-\sum_{k=1}^n P_k$.
\end{theorem}

\textit{Proof.}---To show the sufficiency, we suppose that there exist such  $\{E_1,\cdots,
E_n\}$, define a POVM $\Pi=(\Pi_1,\cdots, \Pi_n)$ as follows:
$\Pi_{k}=P_k+E_k$ for each $1\leq k\leq n$. It is easy to verify
that $\Pi$ is a PPT measurement that perfectly discriminates
$\mathcal{S}$.

Now we turn to show the necessity. Suppose $\mathcal{S}$ is
perfectly distinguishable by some PPT POVM, say
$(\Pi_1,\cdots,\Pi_n)$. Take $E_k=\Pi_k-P_k$ for each $1\leq k\leq
n$. Then $\sum_{k=1}^n E_k=P_0$. To complete the proof, it suffices
to show $E_{k}\geq 0$. By the assumption, we have
$\tr(\Pi_k\rho_k)=1$.  Then the positivity of $E_k$ follows directly
from Lemma \ref{tech}. \hfill
$\blacksquare$

Some special but interesting cases of Theorem \ref{exactdis} deserve
careful investigations. When the supports of the states in
$\mathcal{S}$ together span the whole state space, i.e.,
$supp(\sum_{k=1}^n\rho_k)=\X\otimes\Y$, $\mathcal{S}$ is perfectly
distinguishable by PPT POVMs if and only if $P_k$ is
PPT for each $1\leq k\leq n$. In particular, an orthonormal
basis of $\X\otimes\Y$ is perfectly distinguishable by PPT POVMs if and only if it is a product basis. This coincides with the case of discrimination by separable POVMs.

The following nice result was proved in \cite{HLVC00}.
\begin{lemma}\label{PPTsep}\upshape
Consider a quantum state $\rho\in \mathcal{L}(\X\otimes \Y)$ with $\mathrm{rank}(\rho)\leq \max\{\dim \X,\dim\Y\}$. Then $\rho$ is separable if and only if it is PPT.
\end{lemma}

Combining the above lemma with Theorem \ref{exactdis}, we can establish the equivalence between distinguishing many pure states by PPT POVMs and by separable POVMs.
\begin{cor}\label{pptdissep}\upshape
Let $\mathcal{S}=\{\psi_1,\cdots,\psi_{D-1}\}$ be a collection of
orthogonal pure quantum states of $\X\otimes \Y$, where $D=\dim\X\dim\Y$. Then $\mathcal{S}$ is
perfectly distinguishable by PPT POVMs if and only if it can be distinguished by separable POVMs.
\end{cor}
\textit{Proof.}---Suppose $\psi_{D}$ be the pure state orthogonal to all elements of $\mathcal{S}$, i.e., $\psi_{D}\psi_{k}=0$ for any $1\leq k\leq D-1$. According to Theorem \ref{exactdis}, we know that $\mathcal{S}$ is PPT distinguishable if and only if
there exist $n$ nonnegative numbers $\{\lambda_1,\cdots,
\lambda_{D-1}\}$ with $\sum_{k=1}^{D-1}\lambda_k=1$ such that $\psi_k+\lambda_k\psi_{D}$ is PPT for each $1\leq k\leq {D-1}$.
Note that the rank of $\psi_k+\lambda_k\psi_{D}$ is at most 2. Invoking Lemma \ref{PPTsep}, we know that $\psi_k+\lambda_k\psi_{D}$ is PPT is and only if $\psi_k+\lambda_k\psi_{D}$ is separable. Thus, $\mathcal{S}$ is PPT distinguishable if and only if $\mathcal{S}$ is separable distinguishable.
\hfill
$\blacksquare$

Also we have the following result.
\begin{theorem}\label{indissubspace}\upshape
Let $\ket{\Phi}$ be an entangled pure state on $\X\otimes\Y$. Then
$\{\ket{\Phi}\}^{\perp}$ has no orthonormal basis perfectly
distinguishable by PPT measurements if and only if
$Sch(\Phi)>2$, where $Sch(\Phi)$ denotes the Schmidt number of $\ket{\Phi}$. In particular, when $Sch(\Phi)=2$,
there always exists an orthonormal basis $\mathcal{B}$ of
$\{\ket{\Phi}\}^{\perp}$ that is perfectly distinguishable by LOCC.
\end{theorem}

\section{A limitation of the HSSH method}
In \cite{HSSH03}, Horodecki \textit{et.al} provided a powerful method allowing
for efficient detection of indistinguishability of orthogonal states via LOCC. Their method, called the HSSH method, is described as follows:
 \begin{itemize}
\item[(1)] Given the states $\{\ket{\psi_k}_{\X_1\Y_1}\}_{k=1}^n\subset \X_1\otimes\Y_1$ to be distinguished,
one chooses $n$ entangled states (detectors) $\{\ket{\phi_k}_{\X_2\Y_2}\}_{k=1}^n\subset \X_2\otimes\Y_2$ and probability distribution $\{p_k\}_{k=1}^n$,
and
constructs a pure state
\begin{equation*}
\ket{\psi}_{\X_1\X_2\Y_1\Y_2}=\sum_k \sqrt{p_k}\ket{\psi_{k}}_{\X_1\Y_1}\ket{\phi_{k}}_{\X_2\Y_2},
\end{equation*}

If Alice ($\X_1$) and Bob ($\Y_1$) are able to distinguish between the states $\{\ket{\psi_k}_{\X_1\Y_1}\}_{k=1}^n$ by LOCC,
they can tell the result of their measurement to Claire ($\X_2$) and Danny ($\Y_2$), who
then share states $\{\ket{\phi_k}_{\X_2\Y_2}\}_{k=1}^n$ with probability $\{p_k\}_{k=1}^n$.

\item[(2)]Applying entanglement transformation criterion \cite{NIE99,JP99}, check if the following transition is possible
(in $\X_1\X_2:\Y_1\Y_2$), $i.e.$, whether the
vector $\sum_kp_k\lambda_k$ majorizes $\lambda$, where $\lambda$
and $\lambda_k$ are vectors of the Schmidt coefficients
of $\ket{\psi}$ and $\ket{\phi_k}$ respectively,
\begin{equation*}
\ket{\psi}_{\X_1\X_2\Y_1\Y_2}\overset{\underset{\mathcal{LOCC}}{}}{\longrightarrow} \{p_k,\ket{\phi_k}_{\X_2\Y_2}\}
\end{equation*}
If the transition is impossible, one can conclude that
the set of orthogonal states \(\left\{\ket{\psi_i}\right\}_{k=1}^{n}\) are indistinguishable by LOCC.
\end{itemize}
The authors raised an open problem in \cite{HSSH03}: whether the HSSH method gives a \textquotedblleft if and only if'' criterion. In other words,
given an ensemble, is it true that they are indistinguishable by LOCC if and only if the HSSH method can detect indistinguishability of the ensemble?

This problem was not answered primarily due to the fact that the mathematical structure of LOCC POVMs is complicated. As a direct consequence, we do not even know how to verify that whether a general set of qutrit-qutrit states are locally distinguishable. The second reason is that HSSH method is quite general and there are too many parameters since the statement is ``there exist a set of entangled states (detectors) $\ket{\phi_k}_{\X_2\Y_2}$ and probability distribution $p_k$ such that the LOCC transition is impossible". Therefore, to show this method is not complete, we would need to show that for any detectors and probability distribution, their method does not work. Fortunately, ``Entanglement Discrimination Catalysis" phenomenon can help us to capture these difficulties. The next theorem gives a negative answer to this problem.
\begin{theorem}
There are locally indistinguishable quantum states which cannot be detected by applying the HSSH method with any detectors.
\end{theorem}
\textit{Proof.}---We use the main result of \cite{YDY11b} to show the validity of this theorem:

Let $\ket{\Psi_k}$ denote the standard Bell states, $$\ket{\Psi_k}=(I_2\otimes \sigma_k)\frac{1}{\sqrt{2}}(\ket{00}+\ket{11}),$$ where $\sigma_k$s are the Pauli matrices given by $\sigma_0=I_2$ and
\begin{eqnarray*}
\sigma_1=\left(
\begin{array}{cc}
 1 & 0  \\
 0 & -1 \\
\end{array}
\right),
\sigma_2=\left(
\begin{array}{cc}
 0 & 1  \\
 1 & 0 \\
\end{array}
\right),
\sigma_3=\left(
\begin{array}{cc}
 0 & -i  \\
 i & 0 \\
\end{array}
\right).
\end{eqnarray*}
In \cite{YDY11b}, we showed that $\mathcal{S}=\{\ket{\chi_i}_{\X\Y}:0\leq i\leq 3\}\subset\X\otimes \Y$ cannot be distinguished by any PPT POVM with $\X=\X_1\otimes \X_2$ and $\Y=\Y_1\otimes \Y_2$, where $\X_1,\X_2$, $\Y_1$, $\Y_2$ are all the two-dimensional Hilbert space and
\begin{align*}
\ket{\chi_0}_{\X\Y}=& ~\ket{\Psi_0}_{\X_1\Y_1}\otimes\ket{\Psi_0}_{\X_2\Y_2},\\
\ket{\chi_1}_{\X\Y}=& ~\ket{\Psi_1}_{\X_1\Y_1}\otimes\ket{\Psi_1}_{\X_2\Y_2},\\
\ket{\chi_2}_{\X\Y}=& ~\ket{\Psi_2}_{\X_1\Y_1}\otimes\ket{\Psi_1}_{\X_2\Y_2},\\
\ket{\chi_3}_{\X\Y}=& ~\ket{\Psi_3}_{\X_1\Y_1}\otimes\ket{\Psi_1}_{\X_2\Y_2}.
\end{align*}
Due to the special structure of $\mathcal{S}$, we further observe a quite surprising ``Entanglement Discrimination Catalysis" phenomenon happening on $\mathcal{S}$. More precisely, with a two-qubit maximally entangled state as resource, says $\ket{\Psi_0}$, we can distinguish among the members of $\mathcal{S}$ locally, and after the discrimination, we are still left with an intact copy of $\ket{\Psi_0}$.

Now we show that the indistinguishability of $\mathcal{S}$ can never be detected by HSSH method, i.e., for any four entangled states $\ket{\phi_k}_{\X_3\Y_3}$ of the $\X_3\Y_3$ system and probability distribution $p_k$, the transformation
$\ket{\psi}_{\X\X_3\Y\Y_3}\overset{\underset{\mathcal{LOCC}}{}}{\longrightarrow} \{p_k,\ket{\phi_k}_{\X_3\Y_3}\}$ can always be accomplished by LOCC
(in $\X\X_3:\Y\Y_3$), where
\begin{equation*}
\ket{\psi}_{\X\X_3\Y\Y_3}=\sum_k \sqrt{p_k}\ket{\chi_{k}}_{\X\Y}\ket{\phi_{k}}_{\X_3\Y_3}.
\end{equation*}
According to ``Entanglement Discrimination Catalysis", we know that $\ket{\psi}_{\X\X_3\Y\Y_3}\ket{\Psi_0}_{\X_4\Y_4}\overset{\underset{\mathcal{LOCC}}{}}{\longrightarrow} \{p_k,\ket{\phi_k}_{\X_3\Y_3}\ket{\Psi_0}_{\X_4\Y_4}\}$ is possible
(in $\X\X_3\X_4:\Y\Y_3\Y_4$).

Notice that a necessary and sufficient condition for the transformation
from a pure state $\ket{\phi}$ to an ensemble of pure states $\{p_k,\ket{\phi_k}\}$ was given in \cite{JP99}.
Namely, let $\lambda$
and $\lambda_k$  be vectors of the Schmidt coefficients
of $\ket{\phi}$ and $\ket{\phi_k}$ respectively.
Then the LOCC transition $\ket{\phi}\to \{p_k,\phi_k\}$ is possible if and only if the
vector $\sum_kp_k\lambda_k$ majorizes $\lambda$.

Apply the above criterion on the entanglement transformation $\ket{\psi}_{\X\X_3\Y\Y_3}\ket{\Psi_0}_{\X_4\Y_4}\overset{\underset{\mathcal{LOCC}}{}}{\longrightarrow} \{p_k,\ket{\phi_k}_{\X_3\Y_3}\ket{\Psi_0}_{\X_4\Y_4}\}$ in (in $\X\X_3\X_4:\Y\Y_3\Y_4$). According to the fact that $\ket{\Psi_0}$ is maximally entangled, we can directly obtain that this criterion also satisfied for transformation (in $\X\X_3:\Y\Y_3$), \begin{equation*}\ket{\psi}_{\X\X_3\Y\Y_3}\overset{\underset{\mathcal{LOCC}}{}}{\longrightarrow} \{p_k,\ket{\phi_k}_{\X_3\Y_3}\}.\end{equation*} Therefore, this entanglement transformation can be accomplished by LOCC.
Thus, the HSSH method can not detect the indistinguishability of $\mathcal{S}$.
\hfill
$\blacksquare$

\section{Indistinguishability of maximally entangled state and its orthogonal complement with arbitrary copies}
It is well-known that local measurements on a composite quantum system,
prepared in one of many mutually orthogonal states, may not reveal in which state the system was
prepared.
In the many copy limit, this kind of nonlocality is fundamentally different for
pure and mixed quantum states \cite{DUAN06, BAN11}. In particular, two orthogonal mixed states that are not distinguishable by local operations and classical communication were discovered, no matter how many copies are supplied, whereas any set of $N$
orthogonal pure states can be perfectly distinguished with $N-1$ copies \cite{WSHV00}. Thus, mixed quantum
states can exhibit a new kind of nonlocality absent in pure states. The main tool used in \cite{BAN11} is a well known result, first proved in \cite{DMSST03}, that the tensor of two UPBs (unextendible
product basis) is still a UPB. In this section, we present two quantum states that are not unambiguously distinguishable by PPT POVMs with an arbitrary number of copies.

Before proving the main result of this section, we first provide the following interesting lemma.
\begin{lemma}\label{copypos}\upshape
Let $A,B\in \mathcal{L}(\X)$ be Hermitian operators such that $A+B$ is positive definite and $A$ is not semi-definite, i.e., $\pm A\ngeqslant 0$. For fixed integer $m$, define $\mathcal{T}=\{A,B\}^{\otimes m}\setminus \{A^{\otimes m},B^{\otimes m}\}$, where the tensor product  of two set $\mathcal{S}_1,\mathcal{S}_2$ is given as $\mathcal{S}_1\otimes \mathcal{S}_2=\{s_1\otimes s_2:s_i\in \mathcal{S}_i, i=1,2\}$. Then there do not exist nonnegative numbers $p_k$ such that
\begin{equation}
A^{\otimes m}+\sum_{T_k\in \mathcal{T}} p_k T_k\geq 0.
\end{equation}
\end{lemma}
\textit{Proof.}---Since $A$ is neither positive semi-definite, nor negative semidefinite, we know that there is nonzero positive semidefinite $Q\in \mathcal{L}(\X)$ such that $\tr(QA)=0$. Thus, $$q:=\tr(QB)=\tr(QA)+\tr(QB)=\tr(Q(A+B))>0.$$
By contradiction, assume that there exist nonnegative numbers $p_k$ such that
\begin{equation}\label{tensorpositive}
A^{\otimes m}+\sum_{T_k\in \mathcal{T}} p_k T_k\geq 0.
\end{equation}
Then we can have
\begin{equation}\label{tensor}
\tr_{1,2\cdots,m-1}[(Q^{\otimes m-1}\otimes I_{\X})(A^{\otimes m}+\sum_{T_k\in \mathcal{T}} p_k T_k)]\geq 0,
\end{equation}
where $\tr_{1,2\cdots,m-1}$ denotes the operation that tracing out the first $m-1$ parties. Eq. (\ref{tensor}) implies that for $T_k=B^{\otimes m-1}\otimes A$, we have $q^{m-1}p_k A\geq 0$ which means that $p_k=0$.

Using the similar technique, we can prove that $p_k=0$ for any $T_k\in\mathcal{T}$.  According to Eq. (\ref{tensorpositive}), we know that $A^{\otimes m}\geq 0$. This is impossible.
\hfill
$\blacksquare$

Now we are ready to present the main result of this section.
\begin{theorem}\label{indiscopy}\upshape
Let $\rho_1=\Phi$ and $\rho_2=(I_{\X\otimes\Y}-\Phi)/(d^2-1)$, where $\ket{\Phi}=\frac{1}{\sqrt{d}}\sum_{k=0}^{d-1}\ket{k}_\X \ket{k}_\Y$ is the standard maximally entangled state on the bipartite system $\X\otimes\Y$ with $d=\dim\X=\dim\Y$. Then for any integer $m$, $\rho_1^{\otimes m}$ and $\rho_2^{\otimes m}$ cannot be distinguished unambiguously by PPT POVMs.
\end{theorem}
\textit{Proof.}---Suppose $\rho_1^{\otimes m}$ and $\rho_2^{\otimes m}$ can be distinguished unambiguously by PPT POVMs, then there is some positive PPT operator $E\in\mathcal{L}(\X^{\otimes m}\otimes\Y^{\otimes m})$ such that $\tr(E\rho_1^{\otimes m})>0$, $\tr(E\rho_2^{\otimes m})=0$.

We can construct another $F\in\mathcal{L}(\X^{\otimes m}\otimes\Y^{\otimes m})$ by
\[
F=\int\limits_{V} VEV^{\dag}dV,
\]
where $V$ ranges over all unitaries $\otimes_{k=1}^m (V_{\X_k}\otimes V_{\Y_k})$ with $V_{\X_k}=V_{\Y_k}^*$, and $V_{\X_k}$ ranges over all unitaries.

According to Lemma \ref{invariant}, we know that $F$ is a
positive PPT operator such that $\tr(F\rho_1^{\otimes m})=\tr(E\rho_1^{\otimes m})>0$, $\tr(F\rho_2^{\otimes m})=\tr(E\rho_2^{\otimes m})=0$, and $F\in span\{P_1,P_2\}^{\otimes m}$, where $P_1$ denotes the projector on $supp(\rho_1)$ and $P_2$ denotes the projector on $supp(\rho_2)$, respectively.
Now there are $p,q$ and $p_k$ such that
\[
F=pP_1^{\otimes m}+\sum_{R_k\in \mathcal{R}} p_k R_k+qP_2^{\otimes m},
\]
where $\mathcal{R}=\{P_1,P_2\}^{\otimes m}\setminus \{P_1^{\otimes m},P_2^{\otimes m}\}$.

One can obtain $p,q,p_k\geq 0$ according to the fact that $F\geq 0$. $\tr(E\rho_1^{\otimes m})>0$ and $\tr(E\rho_2^{\otimes m})=0$ imply  $p>0$ and $q=0$.
Note that $P_1^{\Gamma}$ is not semi-definite, $P_2^{\Gamma}>0$ and $P_1^{\Gamma}+P_2^{\Gamma}=I_{\X^{\otimes m}\otimes \Y^{\otimes m}}>0$. Then
\[
F^{\Gamma}=p(P_1^{\Gamma})^{\otimes m}+\sum_{R_k\in \mathcal{R}} p_k R_k^{\Gamma}.
\]
Lemma \ref{copypos} implies that $F^{\Gamma}/p$ is not positive, i.e., $F^{\Gamma}$ is not positive.

Thus, there is no positive PPT operator $E$ such that $\tr(E\rho_1^{\otimes m})>0$ and $\tr(E\rho_2^{\otimes m})=0$. That is,
$\rho_1^{\otimes m}$ and $\rho_2^{\otimes m}$ cannot be distinguished unambiguously by PPT POVMs.
\hfill $\blacksquare$

%-----------------------------------------------------------------------------%
\section{Entanglement cost of distinguishing quantum states}
\label{sec:Entancost}
%-----------------------------------------------------------------------------%
The entanglement cost of state discrimination by LOCC operations was studied in \cite{COH08}. In general, it is quite difficult to deal with the LOCC (separable) POVMs directly. In order to obtain some information about the entanglement cost of the distinguishability using LOCC, we consider the entanglement cost of distinguishing quantum states by PPT POVMs. The examples we considered are quite simple which enable us to obtain analytical results through the techniques developed in \cite{YDY11b}.

\subsection{Distinguishing three Bell states}
Bell states have very nice symmetric properties and they represent the simplest possible examples of entanglement. Previously it was known that two Bell states are locally distinguishable, and three Bell states are locally indistinguishable \cite{GKRS+01}. Indeed, the indistinguishability of three Bell states remain even under separable operations \cite{DFXY09}.
In this subsection, we study the problem of
entanglement cost of distinguishing three Bell states.

First, we can give a lower bound of entanglement cost
for distinguishing three Bell states by LOCC measurements using the HSSH method \cite{HSSH03}:
Suppose $\{\ket{\Psi_k}_{\X_1\Y_1}\otimes\ket{\beta}_{\X_2\Y_2}:1\leq k\leq 3\}$ can be distinguished locally, where $\ket{\beta}_{\X_2\Y_2}=\sqrt{\lambda_0}\ket{00}+\sqrt{\lambda_2}\ket{11}$ such that $\lambda_0\geq \lambda_1\geq 0$ and $\lambda_0+\lambda_1=1$ is the entanglement resource and $\ket{\Psi_k}$ are Bell states.
Now we can construct
another quantum state $\ket{\varphi}_{\X\Y}$ as
\begin{equation*}
\ket{\varphi}_{\X\Y}=\frac{1}{\sqrt{3}}\sum_{k=1}^3\ket{\Psi_k}_{\X_1\Y_1}\ket{\beta}_{\X_2\Y_2}\ket{\Psi_k}_{\X_3\Y_3},
\end{equation*}
where $\X=\X_1\otimes\X_2\otimes\X_3$ and $\Y=\Y_1\otimes\Y_2\otimes\Y_3$.
Since $\{\ket{\Psi_i}_{\X_1\Y_1}\otimes\ket{\beta}_{\X_2\Y_2}\}$ can be distinguished locally,
we have
\begin{equation*}
\op{\varphi}{\varphi}\overset{\underset{\mathcal{LOCC}}{}}{\longrightarrow}\frac{1}{3}\sum_{k=1}^3\op{k}{k}\otimes\op{\Psi_k}{\Psi_k}\overset{\underset{\mathcal{LOCC}}{}}{\longrightarrow}\op{\Psi_0}{\Psi_0}.
\end{equation*}
According to the condition for entanglement transformation
between bipartite pure states \cite{NIE99}, we can assert that
\begin{equation*}
\frac{3}{4}\lambda_0\leq 1/2\Rightarrow \lambda_0\leq \frac{2}{3}.
\end{equation*}
This argument shows that
$\sqrt{2/3}\ket{00}+\sqrt{1/3}\ket{11}$  is necessary
for distinguishing three Bell states locally.

The above method can be used directly to show that maximally entangled state is needed for distinguishing four Bell states locally. The only remaining case is that how much entanglement is required to distinguishing three Bell states since two Bell states are locally distinguishable.
It is worth noting that no LOCC protocol is known to distinguish three Bell states using $\sqrt{2/3}\ket{00}+\sqrt{1/3}\ket{11}$ as a resource. Therefore a quite interesting problem might be ``is partial entanglement helpful for distinguishing Bell states?"

We can prove that
$\sqrt{2/3}\ket{00}+\sqrt{1/3}\ket{11}$  is both necessary and
sufficient for distinguishing three Bell states by
PPT POVMs.
\begin{theorem}\label{main2}
$\mathcal{T}=\{\ket{\Psi_k}_{\X_1\Y_1}\otimes \ket{\alpha}_{\X_2\Y_2}:1\leq k\leq 3\}$ can be
distinguished by some PPT POVM if and only if $\lambda_0\leq 2/3$, where $\ket{\alpha}=\sum\limits_{i=0}^{n-1}\sqrt{\lambda_i}\ket{ii}$ is normalized with Schmidt coefficients $\lambda_0\geq \lambda_1\geq\cdot\cdot\cdot\geq\lambda_{n-1}\geq 0$.
\end{theorem}
\textit{Proof}---For ease of presentation, we first outline the key
proof ideas for the \textquotedblleft only if" part as follows. We can choose
$(C_k)_{k=1}^3$ from $M_{\mathcal{T}}$, where $M_{\mathcal{T}}$ denotes the set of PPT POVMs
that can distinguish $\mathcal{T}$. One can then construct a new POVM $(\Pi_k)_{k=1}^3\in M_\mathcal{T}$
with a highly symmetrical properties by
exploring the convexity of $M_{\mathcal{T}}$ and symmetries of $\mathcal{T}$. The form of
$(\Pi_k)_{k=1}^3$ enables us to bound $\lambda_0$ by calculating its partial
transpose directly.

We start to describe how to construct the desired
$(\Pi_k)_{k=1}^3$ by noticing the following properties of $M_{\mathcal{T}}$ and
$\mathcal{T}$:

First, $M_{\mathcal{T}}$ is convex, i.e., for any $0\leq\lambda\leq 1$,
$$(C_k)_{k=1}^3,(D_k)_{k=1}^3\in M_{\mathcal{T}}\Rightarrow (\lambda C_k+(1-\lambda)D_k)_{k=1}^3\in M_{\mathcal{T}}.$$

Second, $\mathcal{T}$ enjoys a number of symmetries:

\textbf{S1}. For any Pauli matrix $\sigma$, $\sigma_{\X_1}\otimes\sigma_{\Y_1}$ preserves $\ket{\Psi_k}_{\X_1\Y_1}\otimes \ket{\alpha}_{\X_2\Y_2}$ in the following way:
\begin{equation*}
(\sigma_{\X_1}\otimes\sigma_{\Y_1})\ket{\Psi_k}_{\X_1\Y_1}\otimes \ket{\alpha}_{\X_2\Y_2}=\pm\ket{\Psi_k}_{\X_1\Y_1}\otimes \ket{\alpha}_{\X_2\Y_2}.
\end{equation*}

\textbf{S2}. $W_{\X_1\Y_1}$ rotates $\ket{\Psi_k}_{\X_1\Y_1}\otimes \ket{\alpha}_{\X_2\Y_2}$,
\begin{eqnarray*}
W_{\X_1\Y_1}\ket{\Psi_1}_{\X_1\Y_1}\otimes \ket{\alpha}_{\X_2\Y_2}=\ket{\Psi_2}_{\X_1\Y_1}\otimes \ket{\alpha}_{\X_2\Y_2},\\
W_{\X_1\Y_1}\ket{\Psi_2}_{\X_1\Y_1}\otimes \ket{\alpha}_{\X_2\Y_2}=\ket{\Psi_3}_{\X_1\Y_1}\otimes \ket{\alpha}_{\X_2\Y_2},\\
W_{\X_1\Y_1}\ket{\Psi_3}_{\X_1\Y_1}\otimes \ket{\alpha}_{\X_2\Y_2}=\ket{\Psi_1}_{\X_1\Y_1}\otimes \ket{\alpha}_{\X_2\Y_2},
\end{eqnarray*}
where $W$ is defined as
\begin{eqnarray*}
W=\frac{1}{2}\left(
\begin{array}{cc}
 -i & 1  \\
 -i & -1 \\
\end{array}
\right)\otimes \left(
\begin{array}{cc}
 i & 1  \\
 i & -1 \\
\end{array}
\right).\end{eqnarray*}

\textbf{S3}. For any diagonal unitary $V=v\otimes v^*\in\mathcal{L}(\X_2\Y_2)$, $V$ preserves $\ket{\Psi_k}_{\X_1\Y_1}\otimes \ket{\alpha}_{\X_2\Y_2}$ for $0\leq k\leq 3$,
\begin{equation*}
V\ket{\Psi_k}_{\X_1\Y_1}\otimes \ket{\alpha}_{\X_2\Y_2}=\ket{\Psi_k}_{\X_1\Y_1}\otimes \ket{\alpha}_{\X_2\Y_2}.
\end{equation*}
Noticing that a local unitary does not change the positivity
of partial transpose, we can construct a POVM $(\Pi_k)_{k=1}^3\in M_T$ by the convexity of $M_T$
and \textbf{S1-S3} such that
\begin{equation}\label{goal1}
\Pi_{k+1}=W_{\X_1\Y_1}\Pi_kW_{\X_1\Y_1}^{\dag}
\end{equation}
for $k=1,2$,
and for $V=v\otimes v^*\in\mathcal{L}(\X_2\Y_2)$,
\begin{equation}\label{goal2}
\Pi_k=V\Pi_kV^{\dag}=(\sigma_{\X_1}\otimes\sigma_{\Y_1})\Pi_k(\sigma_{\X_1}\otimes\sigma_{\Y_1}).
\end{equation}
Eqs. (\ref{goal1}) and (\ref{goal2}) have greatly restricted the form of $(\Pi_k)_{k=1}^3$.
So, we shall obtain the required $(\Pi_k)_{k=1}^3$ from any POVM $(C_k)_{k=1}^3\in M_{\mathcal{T}}$ by the following three relatively simpler steps:

\textbf{Step 1}: Notice that for a Pauli matrix $\sigma$, we have
\begin{eqnarray*}
((\sigma_{\X_1}\otimes\sigma_{\Y_1})C_k(\sigma_{\X_1}\otimes\sigma_{\Y_1}))_{k=1}^3\in M_{\mathcal{T}}
\end{eqnarray*}
Invoking \textbf{S1} and the convexity of $M_{\mathcal{T}}$, we know that
\begin{eqnarray*}
(D_k)_{k=1}^3=(\frac{\sum_{\sigma}(\sigma_{\X_1}\otimes\sigma_{\Y_1})C_k(\sigma_{\X_1}\otimes\sigma_{\Y_1})}{4})_{k=1}^3\in M_{\mathcal{T}},
\end{eqnarray*}
and each measurement operator $D_k$ is of the form $\sum_j \Psi_j\otimes D^{(kj)}$ for $1\leq k\leq 3$ by noticing that $\sum_{i=0}^3 (\sigma_i\otimes\sigma_i) M(\sigma_i\otimes\sigma_i)$ is diagonal under Bell basis for any $4-$dimensional matrix
$M$.

\textbf{Step 2}: According to \textbf{S2}, one can verify that
\begin{eqnarray*}
(F_k)_{k=1}^3&=&W_{\X_1\Y_1}(D_3,D_1,D_2)W_{\X_1\Y_1}^{\dag}\in M_{\mathcal{T}},\\
(G_k)_{k=1}^3&=&W_{\X_1\Y_1}^{\dag}(D_2,D_3,D_1)W_{\X_1\Y_1}\in M_{\mathcal{T}}.
\end{eqnarray*}
Invoking the convexity of $M_{\mathcal{T}}$ again, we have
\begin{equation*}
(J_k)_{k=1}^3=(\frac{D_k+F_k+G_k}{3})_{k=1}^3\in M_{\mathcal{T}}.
\end{equation*}
We know that for $k=1,2$,
\begin{equation*}
J_{k+1}=W_{\X_1\Y_1}J_kW_{\X_1\Y_1}^{\dag}.
\end{equation*}

\textbf{Step 3}: Invoking \textbf{S3}, we obtain that for any diagonal unitary $V=v\otimes v^*\in\mathcal{L}(\X_2\otimes\Y_2)$,
\begin{equation*}
(L_k)_{k=1}^3=(VJ_kV^{\dag})_{k=1}^3\in M_{\mathcal{T}}.
\end{equation*}
Then we know that
\begin{equation*}
(\Pi_k)_{k=1}^3=(\int\limits_{V}VJ_kV^{\dag}dV)_{k=1}^3\in M_{\mathcal{T}},
\end{equation*}
where $V$ ranges over all diagonal unitaries of form $v\otimes v^*$. One can readily verify that $(\Pi_k)_{k=1}^3$ satisfies Eqs. (\ref{goal1}) and (\ref{goal2}).

Without loss of generality, assume that
\begin{equation*}
\Pi_1=\sum_{ij}N^{(ij)}\otimes \op{ij}{ij}+\sum_{i\neq j}R^{(ij)}\otimes\op{ii}{jj}),
\end{equation*}
where $N^{(ij)},R^{(ij)}\in\mathcal{L}(\X_1\otimes \Y_1)$ are Hermitian with eigenvectors $\ket{\Psi_k}$.
Let
\begin{eqnarray*}
N^{(ij)}=a_{ij}\Psi_0+b_{ij}\Psi_1+c_{ij}\Psi_2+d_{ij}\Psi_3,\\
R^{(ij)}=e_{ij}\Psi_0+f_{ij}\Psi_1+g_{ij}\Psi_2+h_{ij}\Psi_3.
\end{eqnarray*}
According to
\begin{equation*}
\Pi_1+W_{\X_1\Y_1}\Pi_1W_{\X_1\Y_1}^{\dag}+W_{\X_1\Y_1}^{\dag}\Pi_1W_{\X_1\Y_1}=I_{\X_1\X_2\Y_1\Y_2},
\end{equation*}
one can conclude that
\begin{equation*}
a_{00}=1/3,b_{00}+c_{00}+d_{00}=1.
\end{equation*}
From $\Pi_1^{\Gamma}\geq 0$, we know that $N^{{(00)}^{\Gamma}}\geq 0$, then $b_{00}\leq 2/3$.
Invoking Lemma 2, we have
\begin{eqnarray*}
\Pi_1\ket{\Psi_1}\otimes\ket{\alpha}=\ket{\Psi_1}\otimes\ket{\alpha}\\
\Rightarrow \sum_{ij}(b_{ij}\op{ii}{ii}+f_{ij}\op{ii}{jj})\ket{\alpha}=\ket{\alpha}.
\end{eqnarray*}
Now we can make the following assertion:
$\ket{e}=\sum\limits_{i=0}^{n-1}\sqrt{\lambda_i}\ket{i}$ is an eigenvector corresponding to
eigenvalue 1 of non-negative matrix $M_A=(x_{ij})$, where $x_{ii}=b_{ii}$ and $x_{ij}=f_{ij}$ for $i\neq j$. The non-negativity of $M_A$ is derived from the fact that
$\Pi_1$ is semi-definite.
Then we obtain $\lambda_0\leq b_{00}\leq 2/3$ by noticing $M_A-\op{e}{e}\geq 0$, and this ends the proof of the \textquotedblleft only if" part.

The proof of the \textquotedblleft if" part is accomplished by giving
the construction of some PPT POVM $(\Pi_k)_{k=1}^3$
which can
distinguish $\mathcal{T}$ with $\ket{\alpha}=\sqrt{2/3}\ket{00}+\sqrt{1/3}\ket{11}$. Put
\begin{eqnarray}
\Pi_1=\left(
\begin{array}{cccc}
N^{(00)} & 0 & 0& R \\
 0 & N^{(01)} & 0 & 0 \\
 0 & 0 & N^{(10)} & 0\\
R & 0 & 0 & N^{(11)}
\end{array}
\right),
\end{eqnarray}
where $N^{(00)},N^{(01)},N^{(10)},N^{(11)},R\in\mathcal{L}(\X_1\Y_1)$ with
\begin{eqnarray*}
N^{(00)}&=&1/3\Psi_0+2/3\Psi_1+1/6\Psi_2+1/6\Psi_3,\\
N^{(11)}&=&1/3\Psi_0+1/3\Psi_0+1/3\Psi_2+1/3\Psi_3=I/3,\\
N^{(01)}&=&N^{10}=1/3\Psi_0+1/6\Psi_1+5/12\Psi_2+5/12\Psi_3,\\
R&=&\sqrt{2}/3\Psi_1-\sqrt{2}/6\Psi_2-\sqrt{2}/6\Psi_3.
\end{eqnarray*}
It is easy to verify that $(\Pi_1,\Pi_2=W_{\X_1\Y_1}\Pi_1W^{\dag}_{\X_1\Y_1},\Pi_3=W_{\X_1\Y_1}^{\dag}\Pi_1W_{\X_1\Y_1})$ is a PPT POVM which can distinguish $\{\ket{\Psi_k}_{\X_1\Y_1}\otimes \ket{\alpha}_{\X_2\Y_2}:1\leq i\leq 3\}$. \hfill $\blacksquare$

\subsection{Distinguishing a pure state and its orthogonal complement}
In this subsection, we consider the entanglement cost
of distinguishing a pure state and its orthogonal complement. Previously, it is known that two pure orthogonal quantum states can always be distinguished locally\cite{WSHV00}, but the statement is not valid for one pure state and its orthogonal complement. It is quite interesting to study the distinguishability of this set of two states since it forms the simplest indistinguishable states in some sense. Luckily, they can reveal some sharp different difference between the discrimination powers of separable POVMs and PPT POVMs.

Let $\rho_1=\op{\Psi}{\Psi}\in \mathcal{L}(\X_1\otimes\Y_1)$ be a pure state with Schmidt number $d$ and $\rho_2=(I-\rho_1)/(\dim\X_1\dim\Y_1-1)$, where $\ket{\Psi}=\sum_{k=0}^{d-1}\sqrt{\lambda_k}\ket{kk}$ with $\lambda_k\geq \lambda_{k+1}$ for all $0\leq k\leq d-2$.

\subsubsection{By Separable POVMs}

The following theorem gives a lower bound of the entanglement
cost of distinguishing a pure state and its orthogonal
complement by separable measurements.
\begin{theorem}
If $\rho_1\otimes\alpha$ and $\rho_2\otimes\alpha$ can be distinguished by separable POVMs unambiguously, then $Sch(\alpha)\geq d$.
\end{theorem}
\textit{Proof.}---Suppose there is some $\ket{\alpha}\in \X_2\otimes\Y_2$ with Schmidt number $r$ such that $\rho_1\otimes\alpha$ and $\rho_2\otimes\alpha$ can be distinguished by separable POVMs unambiguously, where $\dim\X_2=\dim\Y_2=r$. Without loss of generality, we assume that $\ket{\alpha}=\sum_{i=0}^{r-1}\ket{ii}/\sqrt{r}$.

According to the unambiguous distinguishability condition \cite{CHE04}, there exist two quantum states $\ket{\varphi}\in\X_1\otimes\X_2$ and $\ket{\chi}\in\Y_1\otimes\Y_2$, such that
\[(\rho_1\otimes\alpha)\ket{\varphi\otimes\chi}\neq 0,~~ \mathrm{and} ~~ (\rho_2\otimes\alpha)\ket{\varphi\otimes\chi}= 0.\]
Thus, $\langle\alpha\ket{\varphi\otimes\chi}=c\ket{\Psi}$ for some nonzero $c\in \complex$.
Furthermore, there exist matrices $N_1\in\mathcal{L}(\X_2,\X_1)$ and $N_2\in\mathcal{L}(\Y_2,\Y_1)$ such that $\ket{\varphi}=(I_{\X_1}\otimes N_1)\ket{\Phi}_{\X_1\X_2}$ and $\ket{\chi}=(I_{\Y_1}\otimes N_2)\ket{\Phi}_{\Y_1,\Y_2}$ with $\ket{\Phi}=\sum_{i=0}^{r-1}\ket{ii}/\sqrt{r}$.
Then we have
\begin{eqnarray*}
c\ket{\Psi}=\langle\alpha\ket{\varphi\otimes\chi}&=&\langle\alpha|(N_1\otimes N_2)\ket{\Phi}_{\X_1\X_2}\otimes\ket{\Phi}_{\Y_1,\Y_2}\\&=&(N_1\otimes N_2)\ket{\Phi}_{\X_1,\Y_1}.
\end{eqnarray*}
Compare the Schmidt number, we have
\begin{eqnarray*}
d=Sch(\ket{\Psi})=Sch(c\ket{\Psi})&=&Sch((N_1\otimes N_2)\ket{\Phi}_{\X_2,\Y_2})\\&\leq& Sch(\ket{\Phi}_{\X_2,\Y_2})=r.
\end{eqnarray*}
This ends the proof. \hfill $\blacksquare$

It is not hard to obtain the following
theorem,
\begin{theorem} Let $\ket{\beta}$ be a pure entangled state with $Sch(\beta)<d$. Then
$\rho_1\otimes\beta$ and $\rho_2\otimes\beta$ cannot be distinguished by LOCC measurements unambiguously, but for some finite integer $m$,  $(\rho_1\otimes\beta)^{\otimes m}$ and $(\rho_2\otimes\beta)^{\otimes m}$ can be distinguished perfectly by LOCC measurements.
\end{theorem}
\textit{Proof.}---The first part can be directly obtained by applying Theorem
11. To show $(\rho_1\otimes\beta)^{\otimes m}$ and $(\rho_2\otimes\beta)^{\otimes m}$ can be distinguished perfectly by LOCC measurements for some $m$, we only need to choose sufficient large $m$
such that $\ket{\beta}_{AB}^{\otimes m}$ can be transformed into a $d\otimes d$ maximally
entangled state by LOCC \cite{NIE99}, then distinguish $\rho_1$ and $\rho_2$ by using
teleportation. \hfill $\blacksquare$

Another direct consequence of Theorem 12 is the following:
\begin{cor}
Let $\mathcal{S}=\{\ket{\psi_1},\cdots,\ket{\psi_D} \}$ be an orthonormal
basis of $\X\otimes\Y$ with $D=\dim\X\dim\Y$. Then $\mathcal{S}\otimes\{\ket{\beta}\}$ can be distinguished by separable POVMs unambiguously only if $Sch(\beta)\geq Sch(\psi_k)$ for any $k$.
\end{cor}
\textit{Proof.}---For any $k$, let $\rho_1=\psi_k$ and $\rho_2=(I-\psi_k)/(D-1)$. Since $\mathcal{S}\otimes\{\ket{\beta}\}$ can be distinguished by separable POVMs unambiguously, we can conclude that $\{\rho_1,\rho_2\}\otimes \{\ket{\beta}\}$ is unambiguously distinguishable by separable POVMs. Then Theorem 12 leads us to $Sch(\beta)\geq Sch(\psi_k)$.
\hfill $\blacksquare$

\subsubsection{By PPT POVMs}

We shall see that a two-qubit maximally entangled state
is always enough for distinguishing a pure state and its
orthogonal complement by PPT POVMs. This can be regarded as the fact that PPT POVMs does not always provide a well enough approximation for separable POVMs, even for this quite simple case.

Before proving this result, we first note the following useful lemma.
\begin{lemma} \label{ptpure}
Any eigenvalue of $\rho_1^{\Gamma}$ lies between $-\sqrt{\lambda_0\lambda_1}$ and $\lambda_0$. Moreover, all the eigenvalues of $\rho_1^{\Gamma}$ are $\pm\sqrt{\lambda_i\lambda_j}$ for $i\neq j$ and $\lambda_i$.
\end{lemma}
\textit{Proof.}---It suffices to note that \begin{eqnarray*}
\rho_1^{\Gamma}&&=\sum_{i}\lambda_i\op{ii}{ii}\\&&+\sum_{i> j}\sqrt{\lambda_i\lambda_j}(\frac{\op{ij+ji}{ij+ji}}{2}-\frac{\op{ij-ji}{ij-ji}}{2}).
\end{eqnarray*}\hfill $\blacksquare$

Now we are ready to prove the following:
\begin{theorem}$\rho_1\otimes\alpha$ and $\rho_2\otimes\alpha$ can be distinguished by PPT POVMs, where $\ket{\alpha}=\frac{1}{\sqrt{2}}(\ket{00}+\ket{11}\in\X_2\otimes\Y_2$.
\end{theorem}
\textit{Proof.}---Consider PPT POVM $(\Pi_1,\Pi_2)$ of the following form:
\begin{eqnarray*} \label{form2}
&\Pi_1&=\left(
\begin{array}{cccc}
A & 0 & 0& B \\
 0 & I/2 & 0 & 0 \\
 0 & 0 & I/2 & 0\\
B & 0 & 0 & A
\end{array}
\right),\\
&\Pi_2&=\left(
\begin{array}{cccc}
I-A & 0 & 0& -B \\
 0 & I/2 & 0 & 0 \\
 0 & 0 & I/2 & 0\\
 -B & 0 & 0 & I-A
\end{array}
\right),
\end{eqnarray*}
where $A,B\in\mathcal{L}(\X_1\otimes\Y_1)$, and $I=I_{\X_1\otimes\Y_1}$.

Let $p=\sqrt{\lambda_0\lambda_1}/(1+\sqrt{\lambda_0\lambda_1})$, $q=1/2-p$ and
\begin{equation*}
A=p\Psi+qI,B=(1-p)\Psi-qI.
\end{equation*}
Notice that $0\leq p\leq 1/3,1/6\leq q\leq 1/2$ and $\Pi_1,\Pi_2\geq 0$.\\
We can verify the following
\begin{equation*}
\Pi_k(\rho_k\otimes \alpha)=\rho_k\otimes \alpha.
\end{equation*}
The positivity of $\Pi_k$ comes from
\begin{equation*}
|B|=\frac{1}{2}\Psi+q(I-\Psi)=A\leq I-A.
\end{equation*}
It is clear that $(\Pi_1,\Pi_2)$ is a POVM which can distinguish $\rho_1\otimes\alpha$ and $\rho_2\otimes\alpha$.
The rest part is to show $\Pi_1,\Pi_2$ both enjoys positive partial
transpose. We only need to verify that
\begin{eqnarray*}
I\geq A^{\Gamma}\geq 0 ~~&\mathrm{and}&~~I/2\geq |B^{\Gamma}|,\\
\Leftrightarrow I\geq p\Psi^{\Gamma}+qI\geq 0~~&\mathrm{and}&~~I/2\geq |(1-p)\Psi^{\Gamma}-qI|.
\end{eqnarray*}
Invoking Lemma 14, the largest eigenvalue and smallest
eigenvalue of $p\Psi^{\Gamma}+qI$ satisfy that
\begin{eqnarray*}
1 \geq q + p \geq q + p\lambda_0,\\
q - \sqrt{\lambda_0\lambda_1}p\geq q-\frac{1}{2}p\geq 0.
\end{eqnarray*}
Also, the largest eigenvalue and smallest eigenvalue of $(1-p)\Psi^{\Gamma}-qI$ satisfy that
\begin{eqnarray*}
(1-p)\lambda_0-q\leq 1-p-q=1/2,\\
-q-(1-p)\sqrt{\lambda_0\lambda_1}=-1/2.
\end{eqnarray*}
Thus, $\Pi_1,\Pi_2$ is a PPT POVM.
\hfill $\blacksquare$

The next result shows that one can always find a partial
entangled state to accomplish the task of distinguishing a pure
quantum state and its orthogonal complement, provided the
pure state is not a two-qubit maximally entangled state.
\begin{theorem}
Suppose $\ket{\Psi}=\sum_{k=0}^{d-1}\sqrt{\lambda_k}\ket{kk}\in\X_1\otimes\Y_1$ is an entangled state with $\lambda_{0}\geq\lambda_{1}\geq\cdots\geq\lambda_{d-1}>0$ and $r=\sqrt{\lambda_0\lambda_1}<\frac{1}{2}$. Then $\rho_1\otimes\alpha$ and $\rho_2\otimes\alpha$ can be distinguished by PPT POVMs for some partial entangled state $\ket{\alpha}=\sqrt{\iota}\ket{00}+\sqrt{1-\iota}\ket{11}\in \X_2\otimes\Y_2$ with $\iota<1/2$.
\end{theorem}
\textit{Proof.}---We construct PPT POVM $(\Pi_1,\Pi_2)$, which can distinguish $\rho_1\otimes\alpha$ and $\rho_2\otimes\alpha$, of the following form:
\begin{eqnarray*} \label{form3}
&\Pi_1&=\left(
\begin{array}{cccc}
A & 0 & 0& B \\
 0 & I/2 & 0 & 0 \\
 0 & 0 & I/2 & 0\\
B & 0 & 0 & C
\end{array}
\right),\\
&\Pi_2&=\left(
\begin{array}{cccc}
I-A & 0 & 0& -B \\
 0 & I/2 & 0 & 0 \\
 0 & 0 & I/2 & 0\\
 -B & 0 & 0 & I-C
\end{array}
\right),
\end{eqnarray*}
where $A,B,C\in\mathcal{L}(\X_1\otimes\Y_1)$, and $I=I_{\X_1\otimes\Y_1}$.

Let $t=\sqrt{\frac{1-\iota}{\iota}}\geq 1$, we require that
\begin{eqnarray*}
&\Pi_1(\rho_1\otimes\alpha)=\rho_1\otimes\alpha~~\mathrm{and}~~\Pi_1(\rho_2\otimes\alpha)=0.\\
&\Longrightarrow A+tB=\Psi~~\mathrm{and}~~C+B/t=\Psi.
\end{eqnarray*}
For simplicity, we study the case that $A,B,C$ enjoys very simple form. More precisely,
we try to find some $t>1$ and real numbers $x,y$ such that $(\Pi_1,\Pi_2)$ is
a PPT POVM with
\begin{equation*}
B=x\Psi-yI,A=(1-tx)\Psi+tyI,C=(1-x/t)\Psi+y/tI.
\end{equation*}
To ensure $(\Pi_1,\Pi_2)$ is a POVM, we would need
\begin{eqnarray*}
\Pi_1\geq 0\Leftrightarrow (x-y)(t+1/t)\leq 1,~y\geq 0,\\
\Pi_2\geq 0\Leftrightarrow y(t+1/t)\leq 1,~x-y\geq 0.
\end{eqnarray*}
Invoking Lemma 14,
\begin{eqnarray*}
\Pi_1^{\Gamma}\geq 0\Leftrightarrow ty-(1-tx)r\geq 0,~y/t-(1-x/t)r\geq 0,\\~|x\lambda_0-y|\leq 1/2,~|xr+y|\leq 1/2,\\
\Pi_2^{\Gamma}\geq 0\Leftrightarrow (1-tx)\lambda_0+ty\leq 1,~(1-x/t)\lambda_0+y/t\leq 1,\\~|x\lambda_0-y|\leq 1/2,~|xr+y|\leq 1/2.
\end{eqnarray*}
Our goal is to find real numbers $t>1,x,y$ such that the above inequality holds. In order to do so, we
choose $t=\min\{\sqrt{\frac{1+r}{r}},\frac{1}{2r}\}$, then $t>1$ and $\iota=\frac{1}{t^2+1}$.

We assign values of $x, y$ such that
\begin{equation*}
xr+y=rt\leq\frac{1}{2},~~(x-y)(t+1/t)=1.
\end{equation*}
That is,
\begin{equation*}
x=\frac{rt^3+rt+t}{(r+1)(t^2+1)}, ~y=\frac{rt^3}{(r+1)(t^2+1)}.
\end{equation*}
One can verify that all these inequalities are satisfied by first
noticing that
\begin{eqnarray*}
0\leq y\leq x-y=\frac{t}{t^2+1}\leq 1/2,\\
ty-(1-tx)r=rt^2-r\geq 0,\\
y/t-(1-x/t)r=rt/t-r=0,\\
|x\lambda_0-y|\leq \max\{|y|,|x-y|\}\leq 1/2,\\
(1-tx)\lambda_0+ty\leq \max\{1-tx+ty,ty\}\leq 1,\\
(1-x/t)\lambda_0+y/t\leq \max\{1-x/t+y/t,y/t\}\leq 1.
\end{eqnarray*}
Thus, $(\Pi_1,\Pi_2)$ is a PPT POVM which can distinguish $\rho_1\otimes\alpha$ and $\rho_2\otimes\alpha$ perfectly.
\hfill $\blacksquare$

\subsection{Distinguishing a maximally entangled state and its orthogonal
complement}
Maximally entangled states play a curial role during the development of quantum information theory. As a special case of the problem we studied in the previous subsection, we want to know how much entanglement is required to distinguish a maximally entangled state and its orthogonal
complement by PPT measurements.
Let $\rho_1=\Phi\in\mathcal{L}(\X_1\otimes\Y_1)$ and $\rho_2=(I_{\X_1\otimes\Y_1}-\Phi)/(d^2-1)$, where $\ket{\Phi}=\frac{1}{\sqrt{d}}\sum_{k=0}^{d-1}\ket{kk}$ is the standard maximally entangled state on $\X_1\otimes\Y_1$ with $d=\dim\X_1=\dim\Y_1$.

\begin{theorem} \label{main2}
If $\rho_1\otimes\alpha$ and $\rho_2\otimes\alpha$ can be distinguished by PPT POVMs with $d=2$ and $\ket{\alpha}=\sin\beta\ket{00}+\cos\beta\ket{11}\in \X_2\otimes\Y_2$ with $0\leq\beta\leq \pi/4$, then $\ket{\alpha}=\ket{\Phi}$.
\end{theorem}
\textit{Proof.}---For given $\ket{\alpha}$, let $(M_1,M_2)$ be the PPT POVM which can distinguish $\rho_1\otimes\alpha$ and $\rho_2\otimes\alpha$. One can construct another PPT POVM $(\Pi_1,\Pi_2)$ satisfying the same property, where
\begin{eqnarray*}
\Pi_k=\frac{1}{2}(\int\limits_{V} VM_kV^{\dag}+\int\limits_{V} VM_k^*V^{\dag})dV,
\end{eqnarray*}
where $V$ ranges over all unitaries with form $v_{\X_1}\otimes v^*_{\Y_1}\otimes u_{\X_2}\otimes u^*_{\Y_2}$ for unitary $v$ and diagonal unitary $u$. Direct calculation leads us to the fact that
\begin{eqnarray*} \label{form2}
\Pi_1=\left(
\begin{array}{cccc}
N^{(00)} & 0 & 0& R \\
 0 & N^{(01)} & 0 & 0 \\
 0 & 0 & N^{(10)} & 0\\
R & 0 & 0 & N^{(11)}
\end{array}
\right),
\end{eqnarray*}
where $N^{(ij)},R\in\mathcal{L}(\X_1\otimes\Y_1)$ with $N^{(ij)}=a_{ij}\Phi+b_{ij}(I_{\X_1\Y_1}-\Phi)$, and $R=x\Phi+y(I_{\X_1\Y_1}-\Phi)$ with $a_{ij},b_{ij}\geq 0$, and real numbers $x,y$.

According to $\Pi_1(\rho_1\otimes\alpha)=\rho_1\otimes\alpha$ and $\Pi_1(\rho_2\otimes\alpha)=0$, we know that
\begin{eqnarray*}
a_{00}+x\cot\beta=1,a_{11}+x\tan\beta=1.\\
b_{00}+y\cot\beta=0,b_{11}+y\tan\beta=0.
\end{eqnarray*}
Note that $\Pi_1^{\Gamma}\geq 0$ and $I-\Pi_1^{\Gamma}\geq 0$ implies that $0\leq {N^{(ij)}}^{\Gamma}\leq I_{\X_1\Y_1}$ for $i,j=0,1$. Therefore,
\begin{eqnarray*}
{N^{(ij)}}^{\Gamma}\geq 0\Rightarrow a_{ij}\leq 3b_{ij},\\
{N^{(ij)}}^{\Gamma}\leq I_{\X_1\Y_1}\Rightarrow 3b_{ij}- a_{ij}\leq 2,\\
a_{00}\leq 3b_{00}\Rightarrow 1\leq (x-3y)\cot\beta,\\
a_{11}\leq 3b_{11}\Rightarrow 1\leq (x-3y)\tan\beta.
\end{eqnarray*}
According to the equations obtained above, we see that
\begin{eqnarray*}
1\leq(x-3y)\cot\beta\times(x-3y)\tan\beta=(x-3y)^2.
\end{eqnarray*}
On the other hand, from $\Pi_1^{\Gamma}\geq 0$ and $\Pi_2^{\Gamma}\geq 0$, one can obtain
\begin{eqnarray*}
(3y-x)^2\leq (3b_{10}-a_{10})(3b_{01}-a_{01}),\\
(3y-x)^2\leq (2+a_{10}-3b_{10})(2+a_{01}-3b_{01}).
\end{eqnarray*}
Thus, $(3y-x)^4$ is less than or equal to
\begin{eqnarray*}
(3b_{10}-a_{10})(2+a_{10}-3b_{10})(3b_{01}-a_{01})(2+a_{01}-3b_{01}).
\end{eqnarray*}
Applying the inequality of arithmetic and geometric means, we obtain
\begin{eqnarray*}
(3b_{10}-a_{10})(2+a_{10}-3b_{10})\leq 1,\\
(3b_{01}-a_{01})(2+a_{01}-3b_{01})\leq 1.
\end{eqnarray*}
Therefore,
\begin{eqnarray*}
(x-3y)^4\leq 1\Rightarrow (x-3y)^2\leq 1.
\end{eqnarray*}
Together the inequality,
\begin{eqnarray*}
1\leq =(x-3y)^2
\end{eqnarray*}
we can conclude that
$|x-3y|=1$ and $|\tan\beta|=1.$
That is, $\ket{\alpha}$ is maximally entangled, i.e., $\ket{\alpha}=\ket{\Phi}$.
 \hfill $\blacksquare$

As a direct consequence of Theorem 17, we have the following
interesting corollary for $d=2$.
\begin{cor}
Among all $2\otimes 2$ states, only the maximally entangled state can help to distinguish a two-qubit basis $\{\ket{\varphi_0},\ket{\varphi_1},\ket{\varphi_2},\ket{\varphi_3}\}$ with $\ket{\varphi_0}=\frac{1}{\sqrt{2}}(\ket{00}+\ket{11})$ by LOCC, Separable or PPT POVMs.
\end{cor}

For general $d$, we have the following:
\begin{theorem} For $\rho_1=\Phi\in\X_1\otimes\Y_1$ be the maximally entangled state with $\dim\X_1=\dim\Y_1=d$, $\rho_2=(I-\rho_1)/(d^2-1)$ and $\ket{\alpha}=\sqrt{\iota}\ket{00}+\sqrt{1-\iota}\ket{11}\in \X_2\otimes\Y_2~$ with $$~\iota=\begin{cases} \frac{1}{d+2} &{\rm if}~d\geq 5,\\ \frac{4}{d^2+4} &{\rm otherwise.}\end{cases}$$ Then
$\rho_1\otimes\alpha$ and $\rho_2\otimes\alpha$ can be distinguished by PPT POVMs.
\end{theorem}
\textit{Proof.}---This is a special case of Theorem 16, we use the notations from the proof of Theorem 16 freely. Consider PPT POVM $\{\Pi_1,\Pi_2\}$ constructed in the proof of Theorem 16,
\begin{eqnarray*} \label{form3}
&\Pi_1&=\left(
\begin{array}{cccc}
A & 0 & 0& B \\
 0 & I/2 & 0 & 0 \\
 0 & 0 & I/2 & 0\\
B & 0 & 0 & C
\end{array}
\right),\\
&\Pi_2&=\left(
\begin{array}{cccc}
I-A & 0 & 0& -B \\
 0 & I/2 & 0 & 0 \\
 0 & 0 & I/2 & 0\\
 -B & 0 & 0 & I-C
\end{array}
\right).
\end{eqnarray*}
where $A,B,C$ are with the following form,
\begin{eqnarray*}
B=x\Phi-yI,A=(1-tx)\Phi+tyI,C=(1-x/t)\Phi+y/tI.
\end{eqnarray*}

Notice that $r=\frac{1}{d}$, we have the following two cases:

Case 1: $d\geq 5$, $t=\sqrt{d+1}$, then $\iota=\frac{1}{d+2}$, we choose
\begin{eqnarray*}
x=\frac{2\sqrt{d+1}}{d+2},y=\frac{\sqrt{d+1}}{d+2}
\end{eqnarray*}
Case 2: $2\leq d\leq 4$, $t=\frac{d}{2}$, then $\iota=\frac{4}{d^2+4}$, we choose
\begin{eqnarray*}x=\frac{d(d+2)^2}{2(d+1)(d^2+4)},\ y=\frac{d^3}{2(d+1)(d^2+4)}.\end{eqnarray*}
One can easily verify that $\{\Pi_1,\Pi_2\}$ is a PPT
POVM which can distinguish $\rho_1\otimes\alpha$ and $\rho_2\otimes\alpha$. \hfill $\blacksquare$\\

Based on the above theorem, we know that the entanglement cost of distinguishing a two-qudit
maximally entangled state and its orthogonal complement can
go to $\mathcal{O}(\frac{\log d}{d})$ by PPT POVMs.

\section{Conclusions}

This paper systematically studied the distinguishability of bipartite quantum states by Positive Operator-Valued Measures with positive partial transpose (PPT POVMs).
Several results of \cite{DFXY09} about separable distinguishability were generalized to the case of PPT distinguishability, and an open problem raised in \cite{HSSH03} was negatively answered. It was  proved that maximally entangled
state and its orthogonal complement, no matter how many
copies are supplied, cannot be distinguished by PPT POVMs,
even unambiguously.

The entanglement cost of distinguishing quantum states by PPT POVMs was carefully examined. The cost of discriminating three Bell states was completely figured out: $\sqrt{2/3}\ket{00}+\sqrt{1/3}\ket{11}$ is sufficient and necessary for distinguishing three
Bell states by PPT POVMs. The problem of how much entanglement is needed for distinguishing a $d\otimes d$ pure state and its orthogonal complement was considered. An upper bound of entanglement cost for this problem was derived for separable operations.
We constructed two orthogonal quantum
states which cannot be distinguished unambiguously by separable
POVMs, but finite copies would make them perfectly
distinguishable by LOCC. It was
showed that a two-qubit maximally entangled states is always
enough for discrimination by PPT POVMs, whereas an entangled state with Schmidt number $d$ is
always needed for distinguishing these two states by separable
POVMs. As a special case, the entanglement cost of distinguishing a $d\otimes d$
maximally entangled state and its orthogonal complement
is estimated: for the two-qubit case, the resource must be a
maximally entangled state, but with the increasing of $d$, the
entanglement resource could chosen arbitrarily close to 0. Our results show that PPT POVMs do not always give a well enough approximation of separable POVMs.

There are still several unsolved problems concerning PPT distinguishability. First, it is interesting to clarify the relation between distinguishability by PPT POVMs, separable POVMs, and LOCC POVMs. A more explicit question would be: when PPT POVMs provide a good enough approximation to separable POVMs and LOCC POVMs? For example, in \cite{YDY11b} we showed that four orthogonal ququad-ququad orthogonal maximally entangled states is locally indistinguishable by proving they are PPT indistinguishable. In that case, PPT POVMs form sufficiently good approximations to LOCC POVMs. However, here we observed that that for the case of higher-dimensional spaces, PPT POVMs may behave very differently from LOCC POVMs. Motivated by Theorem 10, we could ask the following interesting question is: whether $\sqrt{2/3}\ket{00}+\sqrt{1/3}\ket{11}$ is sufficient for distinguishing three Bell states
by separable POVMs or LOCC POVMs? Second, the entanglement cost problem of distinguishing a $d\otimes d$ maximally entangled state and its orthogonal complement by separable POVMs (or LOCC POVMs) is of special interest, for instance, is a $d\otimes d$ maximally entangled state always required?
Another problem for further study is to find more applications for PPT distinguishability. In particular, it would be of great interest to obtain some connection between PPT distinguishability to other important concepts in quantum information theory; for instance, we may try to employ PPT distinguishability as a tool to give an upper bound of the environment-assisted classical capacity of quantum channels.
\smallskip

% or


\section*{Acknowledgment}

We are grateful to Dr. Cheng Guo for useful discussions. Part of this work was completed during the N. Yu's visit to Center for Quantum Information Science and Technolog, Tsinghua University.

\ifCLASSOPTIONcaptionsoff
  \newpage
\fi

\end{document}